\documentclass[12pt]{article}
\usepackage{amsmath}
\usepackage{amsfonts}
\usepackage{amssymb}
\usepackage{amstext}
\usepackage{graphicx}
\usepackage{epsfig}%
\usepackage{graphicx}%
\usepackage{color}
\title{\centerline 
\bf  A toy model based analysis on the effect of the Lee-Wick partners
in the evolution of the early universe}
\author{ Kaushik Bhattacharya$^{\$}$,
Suratna Das$^\dagger$
\thanks{email:\,\,\,$^{\$}$kaushikb@iitk.ac.in, $^\dagger$suratna@tifr.res.in,} 
\\
\normalsize
$^{\$}$Department of Physics, Indian Institute of Technology, Kanpur,
\\
\normalsize
Kanpur 208016, India\\
\normalsize
$^\dagger$Tata Institute of Fundamental Research, 
Homi Bhabha Road,\\
\normalsize
Colaba, Mumbai 400005, India}
\begin{document}
\maketitle
\begin{abstract}
In the present article the thermodynamic results of the Lee-Wick
partner infested universe have been applied in a toy model where there
is one Lee-Wick partner to each of the standard model particle and
more over the longitudinal degrees of freedom of the massive partners
of the standard massless gauge bosons are neglected at high
temperatures. For practical purposes, the chiral fermionic sector of
Lee-wick theories requires two Lee-Wick partners per fermion which
opens up the possibility for a negative energy density of the early
universe. A toy Lee-Wick model with one fermionic partner relaxes such
oddities and hence easy to deal with. In a similar way, the
longitudinal degrees of freedom of the massive gauge boson partners
also have the potential to yield negative energy densities and thus
those will be neglected in a toy model study. In such a toy model one
can analytically calculate the time-temperature relation in the very
early radiation dominated universe which shows interesting new
physics. The article also tries to point out how a Lee-Wick particle
dominated early cosmology transforms into the standard cosmological
model. Based on the results of this toy model analysis a brief
discussion on the more realistic model, which can accommodate two
Lee-Wick partners for each standard fermionic field and the
longitudinal degree of freedom of partners of the gauge fields, is
presented. It has been shown that such an universe is mostly very
difficult to attain but there are certain conditions where one can
indeed think of such an universe which can evolve into the standard
cosmological universe in a short time duration.
\end{abstract}

\section{Introduction}

In 1969 Lee and Wick made an attempt \cite{Lee:1969fy,Lee} to
construct a Lorentz invariant, gauge invariant, unitary and
divergence-free Quantum Electrodynamics (QED) by introducing unusual
partners of the normal fields in the Lagrangian.  The partner fields,
called conventionally as the Lee-Wick partners, lived in an indefinite
metric Hilbert space. In recent years the Lee-Wick theory, mainly
constructed for taming divergences in QED, was generalized to
construct a Lee-Wick theory of the standard model of particle physics
\cite{Grinstein}. In the Lee-Wick standard model the authors
constructed a higher-derivative version of quantum field theory and
showed that the higher derivatives in the theory can be removed by
introducing auxiliary Lee-Wick partner fields. It is also shown in
\cite{Grinstein} that such a quantum field theory can be taken as an
extension of the standard model where the mass of the Higgs field is
stable against the quadratically divergent radiative corrections and
thus solves the `Hierarchy puzzle'. Later on various aspects of this
remarkable result have been explored in several works.  Some
interesting applications of the Lee-Wick idea show the scope and depth
of this idea. In \cite{Carone:2008bs} a minimal extension of Lee-Wick
Standard model (LWSM) is considered analyzing its signatures in LHC,
in \cite{Carone1} a Lee-Wick standard model is analyzed where each
standard model particle is accompanied by two Lee-Wick partners, in
\cite{Carone2} gauge-coupling unification is achieved within the
framework of Lee-Wick standard model and in \cite{Carone3,
  Alvarez:2011ah} analysis of two-Higgs doublet models where one of
the doublet contains Lee-Wick fields is done. In \cite{Krauss:2007bz}
the process $gg\rightarrow h_0\rightarrow \gamma\gamma$ is studied in
the framework of LWSM where it has been pointed out that small changes
in the rate of these processes due to presences of Lee-Wick fields can
be treated as a distinguishing feature of LWSM from other models such
as universal extra dimensions. In \cite{Figy:2011yu} Higgs pair
production processes $gg\rightarrow h_0h_0$ and $gg\rightarrow h_0
\tilde p_0$ are studied in LWSM framework to point out that the LW
Higgs can be seen in the LHC upgrade in 2012 particularly when the
new LW Higgs states are below the top pair threshold. The Lee-Wick
field theories described in \cite{Grinstein} has also been extended to
applications in cosmology where in \cite{Cai:2008qw} presence of
Lee-Wick scalar fields in early universe leading to non-singular
bouncing universe is analyzed and in \cite{Karouby:2010wt,
  Karouby:2011wj} the stability condition on presence of radiation
fields and its Lee-Wick partners during such a bouncing universe is
studied. Scalar perturbations in the Lee-Wick bouncing universe
\cite{Cai:2008qw} have been studied choosing spatially flat gauge in
\cite{Cho:2011re} and the corresponding power spectrum has also been
analyzed.

In this work we will mainly focus on the main results of \cite{Fornal}
and \cite{suratna} where the authors have independently derived the
thermodynamic properties of the Lee-Wick resonances at
high-temperatures i.e.  when the temperature of the fluid containing
both the standard model particles and their Lee-Wick partners is much
higher than the mass of the heavy Lee-Wick resonances. It is shown
that the Lee-Wick fields at high temperature contribute negatively to
the energy density and pressure, although the net pressure and energy
density which gets contributions from the standard model particles and
their Lee-Wick partners remain positive. Two different approaches were
taken in these two works, \cite{Fornal} and \cite{suratna}, to derive
the thermodynamic properties of Lee-Wick resonances at high
temperature. As a functional integral formulation of Lee-Wick theories
is not properly derived, the authors of \cite{Fornal} have followed a
method of statistical field theory developed in \cite{Dashen} to
calculate the thermodynamic properties of these unusual Lee-Wick
partners which are treated as unstable resonances.  On the other hand
the authors of \cite{suratna} work with a variant of Lee-Wick's
original idea, first predicted by Boulware and Gross in
\cite{Boulware:1983vw}. In this formulation the unusual Lee-Wick
fields, unlike Lee-Wick's initial work, live in a definite metric
space but carry negative energy. Using these negative energy fields
one can reproduce the same thermodynamic properties as in
\cite{Fornal} from the first principles of statistical mechanics.
Without dealing much with the intricacies of these two methods, we
will follow the main results of these two works \cite{Fornal,suratna}.
It is to be noted that although, by assumption, an indefinite metric
state cannot be an initial or final state of a scattering process
still they may contribute to the thermodynamics of the system. In this
regard one can envisage the Lee-Wick particles as highly unstable
resonances arising as intermediaries in the scattering or decay
processes of standard model particles as elaborated in
\cite{Fornal}. The resonances live momentarily and can contribute to
the thermodynamics of the system.

Within the standard framework of cosmological evolution, the universe
evolves through different phases such as radiation dominated era (when
the cosmic fluid has a state parameter $\omega=1/3$), matter dominated
era (with $\omega=0$), or inflationary and present dark energy
dominated phases of exponential expansion (with $\omega<-1/3$). On the
other hand a cosmic fluid infested with Lee-Wick resonances at very
high temperatures leads to an unusual equation of state ($\omega=1$)
\cite{Fornal,suratna} which might have affected the evolution of the
early universe in an unconventional way\footnote{Although Lee-Wick
  thermodynamics predict an equation of state where $\omega\sim 1$ it
  has been verified that for the toy model with single LW partner for
  each particle the speed of sound $c_s=\sqrt{dp/d\rho}$ remains
  smaller than one and causality is maintained \cite{Fornal}.}. This
present article deals with features of such an unconventional
cosmological scenario. According to Lee-Wick theories the Lee-Wick
partners of the standard model particles are very short-lived and can
only exist as resonances. The unusual equation of state $\omega=1$
arises when the temperature $(T)$ of the cosmic fluid is high enough
to thermalize the short-lived Lee-Wick resonances (i.e. $T\gg {M}$,
where ${M}$ is the mass of the Lee-Wick resonance). Thus we will deal
with an unconventional radiation dominated era when the temperature of
the universe is high enough so that all the particles, including the
Lee-Wick resonances, are thermalized but the equation of state is not
the standard one corresponding to the radiation dominated phase.

It is to note at this point that the Lee-Wick particles cannot occur
as initial or final states in any scattering or decay process but can
exist momentarily as intermediate states. They are introduced in a
theory to overcome the ultraviolet divergences plaguing it. When the
temperature of the system is much higher than the masses of the
Lee-Wick resonances $(T\gg {M})$ then the normal particles can have
energy-momentum greater than $M$ and consequently the unstable
resonances can actually ``live'' momentarily and get
equilibrated. When the temperature of the universe decreases and $(T <
{M})$ then the energy-momentum of the initial particles undergoing
scattering will be less than $M$ and eventually the Lee-Wick
resonances cannot be produced momentarily, they will only act as
virtual particles looping in the propagators. Thus as the temperature
of the universe becomes less than the mass of a Lee-Wick partner the
ephemeral Lee-Wick resonance will not get a chance to be thermally
equilibrated and practically the Lee-Wick partner will thermally
decouple from the plasma. A Lee-Wick partner decouples from the
equilibrated thermal plasma when it looses its property to be produced
momentarily as an intermediate state in a standard scattering or decay
process and this happens when $T<M$.  When the Lee-Wick partners
decouple they simply act as virtual particles and do not contribute to
energy density and pressure of the universe.  In conventional
cosmology we refer to decoupling of a particle from the cosmic plasma
when their interaction rate ($\Gamma$) becomes less than the expansion
rate ($H$) of the universe i.e. $\Gamma<H$. But as the Lee-Wick
partners can exist only as intermediate states, they do not have such
conventional interactions with the cosmic plasma. Thus the only way
they can decouple from the cosmic soup when $T < {M}$ is by being
`non-thermal' as discussed above. We will show later that this way of
becoming `non-thermal' produces rapid temperature and entropy
fluctuations of the remaining cosmic fluid which will lead the system
to go out of equilibrium momentarily.

The rest of the article is presented in the following way. The next
section deals with the basic thermodynamic parameters of a system
where each standard particle has one Lee-Wick partner and where one
neglects the longitudinal degree of freedom of the massive partners of
the massless standard gauge bosons. This is more like a toy model as
because in a realistic Lee-Wick standard model there are two fermionic
partners for each standard chiral fermion \cite{Wise,Espinosa} and the
longitudinal mode of the massive gauge boson partners cannot be
neglected. We still present two sections devoted to this toy model as
most of the interesting physics which will follow in the realistic
case can be qualitatively understood by the toy model thermodynamics.
Section \ref{cosmprop} deals with the cosmological effects arising out
of the thermodynamics of the Lee-Wick partner infested phase.  The
properties of the universe where each standard boson has one Lee-Wick
partner, each standard fermion has two Lee-Wick partners and the
partners of the gauge bosons have three degrees of freedom are
discussed in section \ref{tlwp}. The final section presents a
discussion about the major observations in the article and it ends
with a brief conclusion.
\section{Thermodynamic properties of the Lee-Wick partner infested universe}
\label{therm}
Lee-Wick partners of the standard model particles do not appear as
initial or final states of any real scattering or decay process.  This
is one of the assumptions of Lee and Wick's proposal \cite{Lee:1969fy,
  Lee}. The Lee-Wick partners can only appear as virtual particles or
resonances. In \cite{Fornal} the authors claimed that the Lee-Wick
resonances, which are unstable, can thermalize if the temperature of
the system is much greater than their masses. The masses of the
Lee-Wick partners are assumed to be much greater than the masses of
the standard model particles.

In the radiation dominated phase of standard cosmology whenever the
temperature of the universe scales down to the mass of a particular
particle species then in general the particle species annihilates by
interacting with its anti-particles. This process can happen in
thermal equilibrium. The analogous picture in the Lee-Wick paradigm
corresponds to the situation where the Lee-Wick resonances
decouple. Later it will be shown that unlike standard particles the
Lee-Wick thermal resonances decouple out of thermal equilibrium.  One
can say at these temperatures the Lee-Wick thermal resonances decouple
from the rest of the plasma which is in thermal
equilibrium. Decoupling starts when the temperature of the universe
becomes equal to the mass of a Lee-Wick resonance. As the masses of
the Lee-Wick partners are supposed to be much higher than their
standard partners the thermal resonances of the Lee-Wick partners will
start to decouple much earlier than their standard model
counterparts. These events can modify the thermodynamics of the very
early universe. These modifications of the thermodynamics of the early
universe can have interesting cosmological impacts.
\subsection{The energy density and entropy density of the Lee-Wick particle 
infested universe}
In the standard model of cosmological evolution the energy density,
pressure and entropy density of ultra-relativistic bosons and fermions
are given as
\begin{eqnarray}
\rho^{(\rm sm)}_b= \frac{g\pi^2 T^4}{30}\,,\,\,\,
p^{(\rm sm)}_b= \frac{g\pi^2 T^4}{90}\,,\,\,\,
s^{(\rm sm)}_b= \frac{2g\pi^2 T^3}{45}\,,
\label{smb}
\end{eqnarray}
and
\begin{eqnarray}
\rho^{(\rm sm)}_f= \frac{7g\pi^2 T^4}{240}\,,\,\,\,
p^{(\rm sm)}_f= \frac{7g\pi^2 T^4}{720}\,,\,\,\,
s^{(\rm sm)}_f= \frac{7g\pi^2 T^3}{180}\,,
\label{smf}
\end{eqnarray} 
where the subscripts $b$ and $f$ stands for bosons and fermions and
$g$ stands for any internal degree of freedom of the
ultra-relativistic species. It has been shown in \cite{Fornal,suratna}
that the energy density and the entropy density of the Lee-Wick
partners for the bosonic and the fermionic cases are given by
\cite{Fornal,suratna}:
\begin{eqnarray}
\rho^{(\rm LW)}_b&=& -g\left(\frac{\pi^2 T^4}{30} - \frac{M^2
    T^2}{24}\right)\,,
\label{lwenb}\\
p^{(\rm LW)}_b&=&-g\left(\frac{\pi^2 T^4}{90} - \frac{M^2
    T^2}{24}\right)\,,
\label{lwpb}\\
s^{(\rm LW)}_b&=&-g\left(\frac{2\pi^2 T^3}{45} - \frac{M^2 T}{12}\right)\,,
\label{lwsb}
\end{eqnarray}
and for the fermionic Lee-wick partners one has
\begin{eqnarray}
\rho^{(\rm LW)}_f&=& -g\left(\frac{7\pi^2 T^4}{240} - \frac{M^2
    T^2}{48}\right)\,,
\label{lwenf}\\
p^{(\rm LW)}_f&=&-g\left(\frac{7 \pi^2 T^4}{720} - \frac{M^2
    T^2}{48}\right)\,,
\label{lwpf}\\
s^{(\rm LW)}_f&=&-g\left(\frac{7\pi^2 T^3}{180} - \frac{M^2 T}{24}\right)\,.
\label{lwsf}
\end{eqnarray}
In the above equations $M$ is the mass of a generic Lee-Wick partner
and as the system is relativistic $T \gg M$. Thus in the toy Lee-Wick
model the net energy density, pressure and entropy density of a
relativistic bosonic field in early universe composed of the standard
model field and its Lee-Wick partner are \cite{Fornal,suratna}:
\begin{eqnarray}
\rho_b&=& \rho^{(\rm sm)}_b + \rho^{(\rm LW)}_b = \frac{g M^2 T^2}{24}\,,
\label{nenb}\\
p_b&=& p^{(\rm sm)}_b + p^{(\rm LW)}_b = \frac{g M^2 T^2}{24}\,,
\label{npb}\\
s_b&=& s^{(\rm sm)}_b + s^{(\rm LW)}_b= \frac{g M^2 T}{12}\,.
\label{nsb}
\end{eqnarray}
In writing the above formulas one assumes the internal degrees of
freedom of the standard model bosons and their Lee-Wick partners are
the same. For the massless gauge bosons like photon and gluons one has
two degrees of freedom whereas their massive Lee-Wick partners seems
to have three. Consequently there can be mismatch of degrees of
freedom producing a negative energy density. Negative energy density
by itself will designate an unstable thermodynamic phase. More over in
cosmology, which stems from an underlying theory of general
relativity, one must satisfy the weak energy condition which states
$\rho \ge 0$. If this condition is not satisfied then the Hubble
parameter turns out to be imaginary (for a spatially flat universe)
and one looses all kinds of predictive power over the evolving cosmos.
The way to avoid the unwanted negative energy condition will be
discussed in section \ref{tlwp}.  In this toy model we will assume
that the effect of this extra longitudinal degree of freedom of the
Lee-Wick partners of the gauge bosons to be negligible at high
temperatures. In the toy model both the massless gauge bosons and
their massive partners will have the same internal degrees of
freedom. The realistic case where one considers the longitudinal
degree of freedom of the massive Lee-Wick partners of the massless
gauge bosons will be taken up in section \ref{tlwp}.

If there is precisely one Lee-Wick partner for each standard model
fermion then the combined energy density, pressure and entropy density
of the ultra-relativistic fermionic field and its partner are given as
\begin{eqnarray}
\rho_f&=& \rho^{(\rm sm)}_f + \rho^{(\rm LW)}_f = \frac{g M^2 T^2}{48}\,,
\label{nenf}\\
p_f&=& p_f^{(\rm sm)} + p_f^{(\rm LW)} = \frac{g M^2 T^2}{48}\,,
\label{npf}\\
s_f&=& s_f^{(\rm sm)} + s_f^{(\rm LW)}= \frac{g M^2 T}{24}\,.
\label{nsf}
\end{eqnarray}
On the other hand, if a normal fermionic field has more than one
unusual partner, which is the case of a more realistic Lee-Wick
standard model \cite{Wise,Espinosa}, then the total energy density can
become negative. The realistic case where each standard fermion has
two Lee-Wick partners will be taken up in the penultimate section of
this article. From the previous two triplets of equations in the
toy Lee-Wick model, one can infer the net energy density and entropy
density of a system of ultra-relativistic standard model particles and
their partners with mass $M_i$, as
\begin{eqnarray}
\rho &=& \frac{T^2}{24}\left[
\sum_{i={\rm bosons}} g_iM_i^2\left(\frac{T_i}{T}\right)^2
+ \frac12\sum_{i={\rm fermions}} g_iM_i^2\left(\frac{T_i}{T}\right)^2
\right]\,,
\label{rho}\\
s &=& \frac{T}{12}\left[\sum_{i={\rm bosons}} g_iM_i^2\left(\frac{T_i}{T}\right)
+ \frac12\sum_{i={\rm fermions}} g_iM_i^2\left(\frac{T_i}{T}\right)
\right]\,,
\label{sent}
\end{eqnarray}
where it has been assumed that each standard model boson (or fermion)
and its Lee-Wick partner are in a thermal equilibrium but all the
different species of particles and their partners may not be sharing
the same temperature. Unlike the energy density and entropy density of
the standard cosmological model the above results do depend upon the
masses of the constituents of the plasma.

If $\tilde{M}$ is the mass of the maximally massive partner field, 
then we can compactly write for the overall energy density as
\begin{eqnarray}
\rho=\frac{\tilde{M}^2}{24}\tilde{g}_*T^2,
\label{energyd}
\end{eqnarray}
where $\tilde{g}_*$ is the effective degree of freedom for energy
calculation and the overall entropy density
\begin{eqnarray}
s=\frac{\tilde{M}^2}{12}\tilde{g}_{*s}T,
\label{entropyd}
\end{eqnarray}
where $\tilde{g}_{*s}$ is the effective degree of freedom for entropy
calculation. The effective degrees of freedom are as
\begin{eqnarray}
\tilde{g}_*= \sum_{i={\rm bosons}} 
g_i\left(\frac{M_i}{\tilde{M}}\right)^2
\left(\frac{T_i}{T}\right)^2
+ \frac12\sum_{i={\rm fermions}} g_i \left(\frac{M_i}{\tilde{M}}\right)^2
\left(\frac{T_i}{T}\right)^2\,,
\label{lwgstar}
\end{eqnarray}
and
\begin{eqnarray} 
\tilde{g}_{*s}= \sum_{i={\rm bosons}} 
g_i\left(\frac{M_i}{\tilde{M}}\right)^2
\left(\frac{T_i}{T}\right)
+ \frac12\sum_{i={\rm fermions}} g_i \left(\frac{M_i}{\tilde{M}}\right)^2
\left(\frac{T_i}{T}\right)\,.
\label{gstars}
\end{eqnarray}
It is seen from these two expressions that if all the standard
particles and their Lee-Wick partners are assumed to be in thermal
equilibrium then the values of $\tilde{g}_{*}$ and $\tilde{g}_{*s}$
are smaller than the corresponding values of $g_{*}$ and $g_{*s}$,
where $g_{*}$ and $g_{*s}$ correspond to the values of the effective
degrees of freedom for an equilibrated plasma in conventional
cosmology.

There are some interesting properties about the effective degrees of
freedom $\tilde{g}_{*}$ and $\tilde{g}_{*s}$ which are worth
noticing. The point which makes the behavior of these effective
degrees of freedom different from their standard cosmological
counterparts, ${g}_{*}$ and ${g}_{*s}$, is the dependence of the
effective degrees of freedom on the masses of the Lee-Wick partners.
In standard cosmological calculations ${g}_{*}$ and ${g}_{*s}$ only
depend upon the internal degrees of freedom of the relativistic
particles constituting the plasma.  In the present case if all the
species of the normal particles and their Lee-Wick partners are in
thermal equilibrium then it can be seen that $\tilde{g}_{*} \ge 1$ and
$\tilde{g}_{*s} \ge 1$. More over if all the species and their
partners are in thermal equilibrium then $\tilde{g}_{*}$ and
$\tilde{g}_{*s}$ do not necessarily decrease as the heaviest Lee-Wick
partner becomes non-thermal and drops out of the energy or entropy
calculations. This may happen because the heaviest Lee-Wick partner's
mass, $\tilde{M}$, appears in the the denominator of the squared mass
fractions in the effective degrees of freedom. If the heaviest partner
becomes non-thermal then naturally the place of $\tilde{M}$ is
replaced by the next heaviest thermalized Lee-Wick partner's mass in
the expressions of the effective degrees of freedom.  As a result of
this the squared mass-fractions appearing in the resulting expressions
of $\tilde{g}_{*}$ and $\tilde{g}_{*s}$ may increase if the mass
difference between the two heaviest Lee-Wick partners is high.
\section{Cosmological effects of the thermodynamic properties of the
  Lee-Wick partner infested universe}
\label{cosmprop}

In the Friedman-Robertson-Walker (FRW) paradigm one can write the
line element as
\begin{eqnarray}
ds^2= dt^2 - a^2(t)d{\bf x}^2\,,
\label{line}
\end{eqnarray}
where $a(t)$ is the scale factor of the expanding universe which is
spatially flat. The Friedman equation for a spatially flat universe
is
%
\begin{eqnarray}
H^2=\frac{8\pi}{3M_{\rm Pl}^2}\rho\,,
\label{hubble}
\end{eqnarray}
where $H=\dot{a}/a$ is the Hubble parameter and the Planck mass
$M_{\rm Pl}$ is related to the gravitational constant as
$G\equiv1/M_{\rm Pl}^2$. The energy density scales with the scale
factor as $\rho(t) \propto a(t)^{-3(1+\omega)}$ where the scale factor
$a(t) \propto t^{\frac{2}{3(1+\omega)}}$. Here $\omega$ is the state
parameter which relates the energy density and pressure for a
barotropic fluid as $p=\omega \rho$. From the results of the last
subsection it can be easily verified that
\begin{eqnarray}
\omega=1\,
\label{olw}
\end{eqnarray}
for a fluid where each ultra-relativistic standard model particle is
accompanied by its relativistic Lee-Wick partner. Consequently for a
radiation dominated universe infested with Lee-Wick particles one must
have
\begin{eqnarray}
\rho(t) &\propto& a(t)^{-6}\,,
\label{endens}\\
a(t) &\propto& t^{\frac13}\,,
\label{ats}
\end{eqnarray}
a result noted earlier in \cite{Fornal}, where in conventional
radiation dominated era $\rho(t)\propto a(t)^{-4}$ and $a(t)\propto
t^{\frac12}$. All of these results point towards a new radiation
dominated era of cosmological evolution which differs from the
standard radiation dominated $(\omega=1/3)$ and matter dominated
$(\omega=0)$ periods.

As in the presence of the thermalized Lee-Wick resonances the entropy
density of the universe is given by Eq.~(\ref{entropyd}), thus for an
isentropic process the temperature of the universe varies with time as
\begin{eqnarray}
T(t)=\frac{T_0}{a^3(t)},
\label{Tt}
\end{eqnarray}
where $T_0$ is a constant. In standard cosmology $T \propto a(t)^{-1}$
for an isentropic process.  Initially the temperature of the universe
is greater than the mass of the heaviest Lee-Wick resonance as a
consequence of which the heaviest Lee-Wick resonance remains in
thermal equilibrium with the rest of the plasma.  In the thermal
Lee-Wick resonance dominated radiation era the temperature decreases
with time till the temperature of the universe becomes such that $T
\sim \tilde{M}$.  At this temperature the Lee-Wick resonance with the
heaviest mass $\tilde{M}$ decouples from the rest of the
plasma. Henceforth the heaviest Lee-Wick partner's mass appearing in
the expressions of energy density and entropy density in
Eq.~(\ref{energyd}) and Eq.~(\ref{entropyd}) will get replaced by the
next heaviest Lee-Wick partner's mass. When $T \sim \tilde{M}$ and the
heaviest Lee-Wick resonance is decoupled, there will be a
discontinuity in various thermodynamic parameters of the system which
will be discussed later.

From the nature of the scaling of temperature, given in the above
equation, it can be inferred that in a universe where the Lee-Wick
partners of the standard model particles are also present the
time-temperature relationship will change. The exact relationship
between cosmic time $t$ and temperature of the universe $T$ is
calculated as follows. From Eq.~(\ref{Tt}) one can immediately write
\begin{eqnarray}
\left(\frac{\dot{T}}{T}\right)=-3H,
\label{tteqn}
\end{eqnarray}
where $H$ is given by Eq.~(\ref{hubble}). Using Eq.~(\ref{hubble})
and Eq.~(\ref{energyd}) one can write the Hubble parameter as
\begin{eqnarray}
H=\sqrt{\pi \tilde{g}_*}\frac{T\tilde{M}}{3M_{\rm Pl}}\,.
\end{eqnarray}
Using the above mentioned value of the Hubble parameter one can solve 
Eq.~(\ref{tteqn}) and get
\begin{eqnarray}
t\sim 10^{-24}(\pi \tilde{g}_*)^{-1/2}
\left(\frac{M_{\rm Pl}}{\tilde{M}}\right)\left(\frac{1{\rm GeV}}
{T}\right)\,{\rm sec}\,.
\label{Ttr}
\end{eqnarray}
The time-temperature relationship for a Lee-Wick infested radiation
dominated universe is distinctly different from the standard
time-temperature relationship of the standard radiation dominated universe
where 
\begin{eqnarray}
t \sim 10^{-6} g_*^{-1/2}\left(\frac{1{\rm GeV}}{T}\right)^2{\rm sec}\,.
\label{Tts}
\end{eqnarray}
The difference between the the two time-temperature relations as
given in Eq.~(\ref{Ttr}) and Eq.~(\ref{Tts}) can have interesting
effects in the cosmological evolution of the very early universe.
From Eq.~(\ref{Ttr}) one can also write
\begin{eqnarray}
T_{\rm GeV}\sim 10^{-24}(\pi \tilde{g}_*)^{-1/2}
\left(\frac{M_{\rm Pl}}{\tilde{M}}\right) \frac{1}{t_{\rm sec}}\,,
\label{ttr}
\end{eqnarray}
where $t_{\rm sec}$ is measured in seconds and $T_{\rm GeV}$ is the
temperature of the system in units of GeV.

In standard cosmology the conventional radiation dominated era sets up
after reheating which `defrosts' the cold, low-entropy universe at the
end of inflationary era. The equilibrated temperature reached after
the reheating of the universe, known as the reheat temperature $T_{\rm
  Rh}$, is considered as the maximum temperature of the following
radiation dominated universe. Considering supersymmetric extension of
standard model of particle physics an upper bound on the reheat
temperature ($T_{\rm Rh}<10^9\sim10^{10}$ GeV) comes from the effects
of overproduction of gravitinos and their decay products on light
element abundances during big bang nucleosynthesis \cite{Ellis}. On
the other hand, the Lee-Wick standard model \cite{Grinstein}, which
successfully addresses the `hierarchy puzzle' of the standard model,
does not require to include supersymmetric partners. Hence in a
cosmological scenario, where the evolution of the universe is
described by Lee-Wick extension of standard model, an upper bound on
reheat temperature can be set by the constraints arising from (i)
generation of Ultra-high energy cosmic rays from super-heavy
long-living $X-$particles during reheating \cite{Kuzmin}, or (ii)
imposing naturalness condition on the self-coupling of the inflaton
$(\lambda_\phi\sim 10^{-13})$ while estimating the reheat temperature
in perturbative reheating scenario (here $T_{\rm Rh}\sim \sqrt{\Gamma
  M_{\rm Pl}}$, $\Gamma$ being the total decay rate of inflaton)
\cite{Greene}. Considering these scenarios, preceded by a generic
inflationary era, we heuristically set the upper bound on reheat
temperature as $T_{\rm Rh}\lesssim 10^{10}$ GeV. A study of
baryogenesis in such a Lee-Wick field infested cosmological scenario
is beyond the scope of this paper and requires further investigation.
With this upper bound on reheat temperature, one can also determine
the age of the universe at the onset of radiation era in standard
cosmology using Eq.~(\ref{Tts}) as
\begin{eqnarray}
t_{\rm rad}\sim 10^{-27}\,\,{\rm sec},
\end{eqnarray}
where we have taken $g_*\sim 100$. 

In a Lee-Wick partner infested cosmology we refer to an era as a
radiation dominated era when all the particles in the cosmic soup are
at thermal equilibrium and are all relativistic. The age of the
universe at the onset of Lee-Wick infested radiation dominated era,
unlike the generic scenario, will depend up on the mass of the
heaviest Lee-Wick partner present in the cosmic soup at that time,
which can be seen from Eq.~(\ref{Ttr}). Taking $\tilde{M}\sim10^9$ GeV
and $\tilde{g}_*\sim80$, the reheat temperature ($T_{\rm Rh}\sim
10^{10}$ GeV) determines the age of the universe at the onset of
the Lee-Wick infested radiation era as
\begin{eqnarray}
t_{\rm rad}^{\rm LW}\sim 10^{-25}\,\, {\rm sec},
\end{eqnarray}
whereas a lower mass of the heaviest Lee-Wick partner, say
$\tilde{M}\sim 10^5$ GeV, delays the onset of a radiation dominated
era where
\begin{eqnarray}
t_{\rm rad}^{\rm LW}\sim 10^{-21}\,\, {\rm sec}.
\end{eqnarray}

Now, we will consider $\tilde{M}\sim10^9$ GeV and $\tilde{g}_*\sim80$
to demonstrate the evolution of a Lee-Wick infested radiation era
since its onset at $t_{\rm rad}^{\rm LW}\sim 10^{-25}$ sec. It can be
seen from Eq.~(\ref{ttr}) that as time progresses the temperature of
the system decreases and at a certain time, when $T \sim \tilde{M}$,
the Lee-Wick partner with the heaviest mass decouples and consequently
its contribution will drop out from the expressions of energy density
and entropy density of the universe. From then onwards the place of
$\tilde{M}$ will be taken up by the mass of the next maximally massive
Lee-Wick partner, which we will take as $\tilde{M}^\prime\sim
3\times10^8$ GeV. As a consequence of this there will be an abrupt
change, specifically a bump in the temperature, when $T \sim
\tilde{M}$.
\begin{figure}[h!]
\centering
\includegraphics[width=10cm,height=15cm,angle=270]{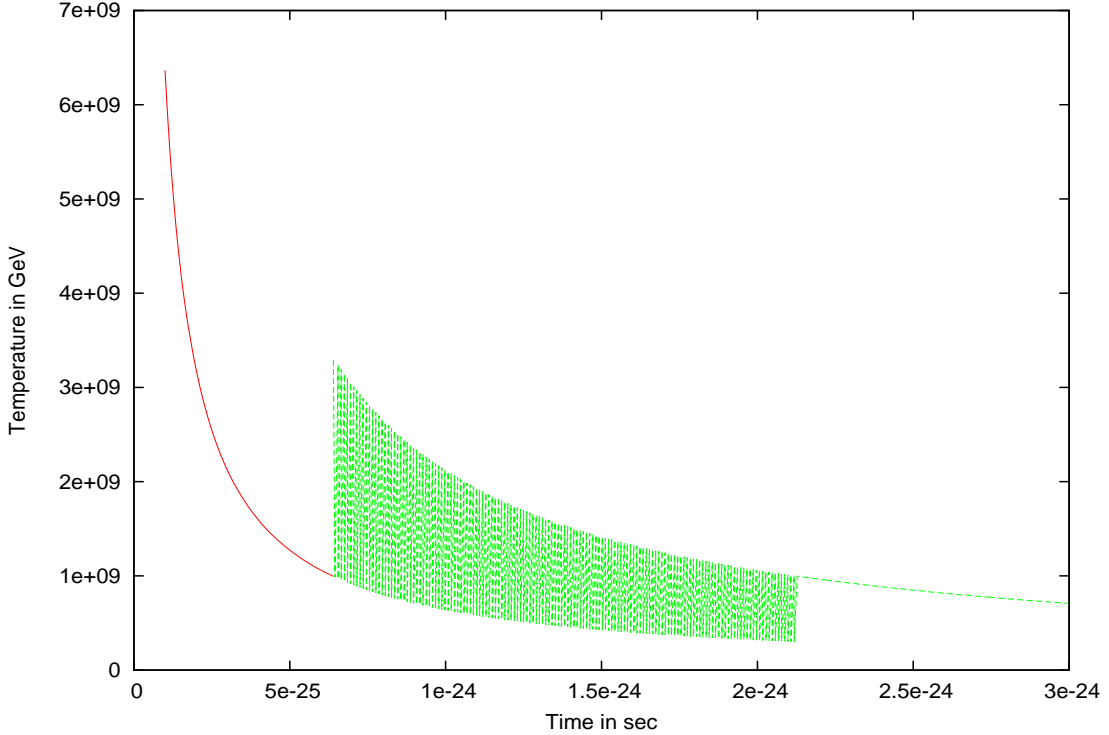}
\caption[]{Figure showing the bump in the temperature as the heaviest
  Lee-Wick partner with mass $10^9$ GeV becomes non-thermal and its
  position is taken up by the next heaviest Lee-Wick partner of mass
  $3\times 10^8$ GeV. For a simple illustration the effective degrees
  of freedom are assumed to remain constant in the process,
  $\tilde{g}_*\sim 80$. The green region in the plot represents a
  region of rapid and violent temperature fluctuations where the
  system has gone out of equilibrium.}
\label{tt:f}
\end{figure}

\begin{figure}[h!]
\centering
\includegraphics[width=10cm,height=15cm,angle=270]{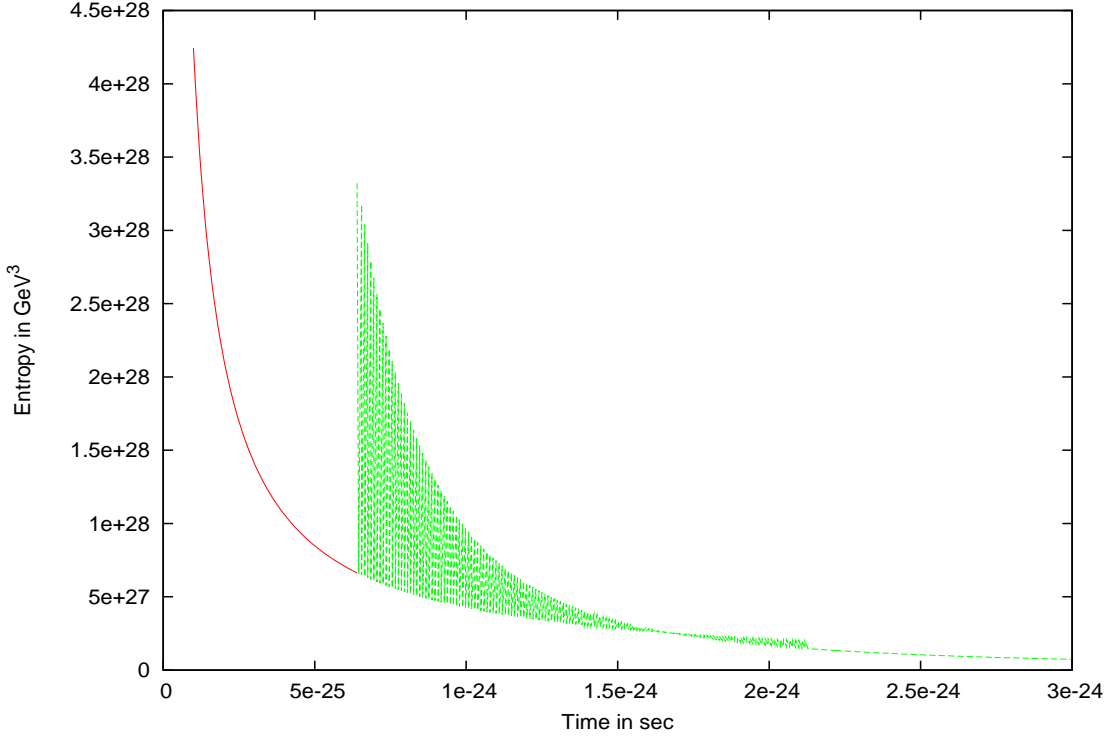}
\caption[]{Figure showing the bump in the entropy density of the
  universe as the heaviest Lee-Wick partner with mass $10^9$ GeV
  becomes non-thermal and its position is taken up by the next
  heaviest Lee-Wick partner of mass $3\times 10^8$ GeV. The effective
  degrees of freedom are assumed to remain constant in the process as
  $\tilde{g}_{*s} \sim 80$. The standard model particle freed of its
  partner is assumed to be a boson with $g=2$. The spiked green region
  in the plot represents a region of rapid and violent entropy
  fluctuations where the system has gone out of equilibrium.}
\label{st:f}
\end{figure}

The discontinuity in the temperature of the universe due to decoupling
of the heaviest Lee-Wick partner from the cosmic soup is shown in
Fig.~\ref{tt:f}. The magnitude of the bump depends upon the difference
of the mass of the maximally massive Lee-Wick partner and the mass of
the second maximally massive Lee-Wick partner. The nature of the
temperature bump, as the thermal resonances decouple, is complex. As
soon as the temperature reaches $\tilde{M}$ there is an abrupt rise in
the temperature which, if greater than $\tilde{M}$, re-thermalizes the
heaviest Lee-Wick mode once again. As the temperature of the
re-thermalized universe tries to come down and the temperature reaches
$\tilde{M}$ again there is a second bump which again re-thermalize the
heaviest Lee-Wick partner. Consequently there are rapid and violent
temperature fluctuations which takes the system out of thermal
equilibrium. The fact that the system goes out of equilibrium can also
be verified by the entropy bump in Fig.~\ref{st:f} where it is shown
that there is an abrupt bump in the entropy of the universe as the
heaviest Lee-Wick resonance decouples.  Equilibrium is regained once
the resultant temperature is smaller than $\tilde{M}$ such that the
maximally massive Lee-Wick partner is not re-thermalized. This effect
is portrayed by the shaded region in Fig.~\ref{tt:f}. The entropy bump
in Fig.~\ref{st:f} can also be explained by an analogous analysis.

It is interesting to note here that as the heaviest Lee-Wick partner
decouples from the cosmic plasma and seizes to be thermalized, it will
leave behind its standard model partner in the cosmic plasma which
being very light will still be thermalized and will contribute to the
energy and entropy density of the radiation dominated universe. Thus
the decoupling of the heaviest of the Lee-Wick partners will change
the expressions of the energy density and entropy density from the
forms they had in Eq.~(\ref{energyd}) and Eq.~(\ref{entropyd}). After
some of the Lee-Wick partners become non-thermal, the form of the
energy density and the entropy density will look like
\begin{eqnarray}
\rho'=\frac{\tilde{M'}^2}{24}\tilde{g'}_*T'^2 + 
\frac{\pi^2 T'^4}{30}\,g_*^\prime\,,
\label{mixede}\\
s'=\frac{\tilde{M'}^2}{12}\tilde{g'}_{*s}T' + 
\frac{2\pi^2 T'^3}{45}\,g_{*s}^\prime\,.
\label{mixeds}
\end{eqnarray}
These expressions show that after the decoupling of few heaviest
Lee-Wick partners the universe will be in a mixed state where the two
components of the mixture have different forms of thermodynamic
parameters. The first terms on the right hand sides of the equations
comes from that part of the plasma which consists of those standard
model particles and their Lee-Wick partners which are still
thermalized at that temperature $T^\prime$. The second term of the
right hand sides of the equations comes from a plasma constituted by
only standard model particles whose Lee-Wick partners have already
decoupled from the cosmic plasma. As the Lee-Wick partners are heavier
than their standard model partners it is safe to assume that these
standard model particles are still relativistic at temperature
$T'$. In the above equations we assume that $\tilde{M'}$ is the mass
of the maximally massive Lee-Wick partner at the corresponding
temperature $T'$. The effective degrees of freedom changes to
$\tilde{g'}_*$ and $\tilde{g'}_{*s}$ from their previous values. Also,
$g_*^\prime$ and $g_{*s}^\prime$ account for the effective degrees of
freedom of those standard model particles whose Lee-Wick partners have
decoupled.

If very few of the Lee-Wick partners have become non-thermal then
$\tilde{g'}_* \gg g_*^\prime$ and $\tilde{g'}_{*s}\gg g_{*s}^\prime$
and more over if the temperature $T'$ happens to be marginally greater
than $\tilde{M'}$, $T' \gtrsim \tilde{M'}$, then the presence of the
Lee-Wick partners predominantly shape the thermodynamics of the
plasma, which can be seen from Eq.~(\ref{mixede}), and consequently
the evolution of the thermodynamic parameters follows the laws as
discussed in this section. In such a case the equation of state of the
plasma will change slightly and $\omega \lesssim 1$.  A simple
situation of this kind is portrayed in the temperature and entropy
bumps in Fig.~\ref{tt:f} and Fig.~\ref{st:f}.  In the figures
$\tilde{g}_*$ and $\tilde{g}_{*s}$ remain approximately constant after
the heaviest Lee-Wick partner become non-thermal and the standard
model particle which looses its partner is a boson with $g=2$.
As more and more Lee-Wick partners become non-thermal we will have
$\tilde{g'}_* \gtrsim g_*^\prime$ and $\tilde{g'}_{*s} \gtrsim
g_{*s}^\prime$, but once $T' \gtrsim \tilde{M'}$ the second terms of
Eq.~(\ref{mixede}) and Eq.~(\ref{mixeds}) start to dominate and
consequently the thermodynamics of the universe transforms from the
unusual Lee-Wick particle dominated phase to the standard radiation
dominated phase where now $\omega \sim 1/3$. The point to be noted is
that the thermodynamic development of the universe can become the
standard one after some of the Lee-Wick partners become non-thermal
and the rest are still there in the plasma. To have the universe
evolve as the standard radiation dominated one with
$\omega\sim\frac13$ one should have the standard model particles,
which are devoid of their Lee-Wick partners, to dominate the energy
density of the universe. From Eq.~(\ref{mixede}) this condition yields
\begin{eqnarray}
 T^\prime>
 0.36\left(\frac{\tilde{g^\prime}_*}{g_*^\prime}\right)^{\frac12}
\tilde{M^\prime}.
\label{T-generic}
\end{eqnarray}
We require the generic radiation dominated era to set well before
neutrino decoupling so that the general scenario of the radiation
dominated universe is not much affected by the presence of the
Lee-Wick partners. Hence we set $t\sim 1$ sec as the time of the onset
of the generic radiation dominated universe with $\omega\sim
\frac13$. Before the onset of the generic radiation domination the
evolution of the universe in governed by the presence of the heavy
Lee-Wick resonances and thus the time-temperature relation will follow
the form given in Eq.~(\ref{ttr}). An upper bound on mass of the
heaviest Lee-Wick partner, which can still remain in the thermal
cosmic soup even after the onset of a generic radiation dominated
universe at $t\sim 1$ sec, thus can be determined by Eq.~(\ref{ttr})
and Eq.~(\ref{T-generic}) as
\begin{eqnarray}
\tilde{M}^\prime < 1.4\times 10^{-2} \,\, {\rm GeV},
\end{eqnarray}
where we have assumed for simplicity $\tilde{g^\prime}_*\approx
g_*^\prime\approx 60$ at the time of onset of generic radiation
domination.
\section{A more realistic approach}
\label{tlwp}

There are some subtle issues related to the number of
fermionic Lee-Wick partners accompanying a normal fermion as
discussed.  The subtlety of the issue is related to how many Lee-Wick
partners does a chiral fermion have?  Initially it was supposed that
there are two but those two cannot be treated as independent degrees
of freedom as they are related to each other in the higher derivative
form of the theory. The main reason for declining the independence of
the two Lee-Wick fermion partners is related to the emergence of
negative energy density. The issue remains subtle as can be seen in
the discussions in \cite{Wise, Espinosa}. In the present section we
will not go into a theoretical debate about this issue, but we will
present a picture of the possible thermodynamics of the universe if
both the fermionic partners of a standard chiral fermion were really
independent degrees of freedom in the auxiliary field approach. It can
be shown that although this scenario is precarious, as it can
potentially lead to an early universe with negative energy density,
there can be some conditions during cosmological reheating which can
evade the difficulties and produce a workable model of early universe
thermodynamics.

More over in the toy model of the Lee-Wick cosmology as discussed in
the sections before we ignored the longitudinal mode of polarization
of the massive Lee-Wick partners of massless vector bosons. In this
section we will like to incorporate their effect in the energy density
and other relevant thermodynamic parameters. If one assumes that both
the Lee-Wick partners of each Standard model fermion are of equal mass
and considerably heavier than their standard model partners, then
considering the bosons and fermions with their Lee-Wick partners
altogether, one can write
\begin{eqnarray}
\rho=\frac{\tilde{M}^2}{24}\tilde{g}_{*N}T^2 - \frac{7\pi^2}{240}
\tilde{g}_FT^4 - \frac{\pi^2}{30}nT^4\,,
\label{energyd2}
\end{eqnarray}
where the new degrees of freedom $\tilde{g}_{*N}$ and $\tilde{g}_F$ are given as
\begin{eqnarray}
\tilde{g}_{*N}= \sum_{i={\rm bosons}} 
g_{iN}\left(\frac{M_i}{\tilde{M}}\right)^2
\left(\frac{T_i}{T}\right)^2
+ \sum_{i={\rm fermions}} g_i \left(\frac{M_i}{\tilde{M}}\right)^2
\left(\frac{T_i}{T}\right)^2\,,
\label{lwgstar2}
\end{eqnarray}
where $g_{iN}$ for bosonic particles stands for the internal degrees
of freedom $g_i$ for the partners of massive standard bosons (may be 2
or 1), while for standard massless vector boson partners it equals
$g_i+1$ where primarily $g_i=2$.  The unpaired fermionic contribution
comes with $\tilde{g}_F$ where
\begin{eqnarray}
\tilde{g}_F = \sum_{i={\rm fermions}} g_i \left(\frac{T_i}{T}\right)^4\,,
\label{lwgf}
\end{eqnarray}
where $\tilde{g}_F$ solely arises from the unpaired fermionic Lee-Wick
partners of the standard model particles. The quantity $n$ is defined
as
\begin{eqnarray}
n= \sum_{i={\rm massive\,vect.\,bosons}} 
\left(\frac{T_i}{T}\right)^4\,.
\label{newn}
\end{eqnarray}
Here $T_i$ is the temperature at which the $i^{\rm th}$ massive
Lee-Wick vector boson partner of a standard massless gauge boson is
equilibrated and $n$ denotes the number of massive vector boson
partners of massless standard gauge bosons if all the species are in
thermal equilibrium at the same temperature $T$. The sum appearing in
Eq.~(\ref{newn}) does not include all the massive Lee-Wick vector
boson partners, it includes only those which are partners of massless
standard gauge bosons.

Assuming all the particle species and their partners are in thermal
equilibrium it can be verified from Eq.~(\ref{energyd2}) that if the
temperature of the universe satisfies the following inequality
\begin{eqnarray}
T \le
\sqrt{\frac{5\,\,\tilde{g}_{*N}}{4\pi^2(n+\frac78\,\tilde{g}_F)}}\,\,\,\, 
\tilde{M}\,,
\label{tcond}
\end{eqnarray} 
then the energy density of the universe can be positive. The
inequality above is very restrictive as for thermal Lee-Wick partners
$T \ge \tilde{M}$. Combining these two inequalities one can say that
the very early universe can have a very small positive energy density
only if
\begin{eqnarray}
\tilde{g}_{*N} \ge \frac{4\pi^2}{5}\left(n+\frac78\tilde{g}_F\right)\,,
\label{ggrel}
\end{eqnarray}
which sets a constraint on the internal degrees of freedom available
for the particles inhabiting the universe during these early
times. Eq.~(\ref{ggrel}) is a difficult one to satisfy as
$\tilde{g}_{*N}$ must be a small number if the Lee-Wick partner masses
have a stiff hierarchy as $\tilde{g}_{*N}$ contains the square of the
ratios of $({M_i}/{\tilde{M}})^2$, where as $\tilde{g}_F$ and $n$ does
not have any such factors. Before we comment on the possibility of any
cosmological phase where one has two Lee-Wick partners of chiral
fermions and considers the longitudinal degrees of freedom of massive
Lee-Wick gauge fields it is better to have an expression of entropy of
such a system. The pressure of such a plasma which consists of two
Lee-Wick partners of one standard chiral fermion is given by
\begin{eqnarray}
p=\frac{\tilde{M}^2}{24}\tilde{g}_{*N}T^2 -
\frac{7\pi^2}{720}\tilde{g}_FT^4 - \frac{\pi^2}{90}nT^4\,.
\label{pres2}
\end{eqnarray}
It must be noted that if the energy density of the Lee-Wick partner
infested universe is positive then the pressure of the same universe
must be positive. 
From the expressions of the
energy density and pressure one can calculate the entropy density of
such a plasma as
\begin{eqnarray}
s=\frac{\tilde{M}^2}{12}\tilde{g}_{*N}T -
\frac{7\pi^2}{180}\tilde{g}_FT^3
- \frac{2\pi^2}{45}nT^3\,.
\label{sdens2}
\end{eqnarray}
From the above expression one can easily see that the entropy
density is not positive definite at high temperatures and if it is
indeed positive it can be very small. If one has to think of a
radiation dominated phase at the time of reheating after inflation one
must have to generate entropy. The Lee-Wick partners of the fermions
and massless gauge bosons does not allow this high entropy generation
during reheating.

In the present scenario one sees that if there are two Lee-Wick
fermionic partners for each standard fermion both energy density and
entropy density remains very low. There can be some interesting cases
where the above difficulties can be circumvented. In the following
discussion we briefly mention some of these conditions :
\begin{enumerate}
\item If the early universe had considerably more number of heavy
  bosons as compared to the total number of chiral fermions and
  massless gauge bosons then the condition set by Eq.~(\ref{ggrel})
  can be satisfied. Such a scenario can be related to the preheating
  scenario \cite{Kofman} where the coherently oscillating classical
  inflaton field decays very rapidly into many heavy bosons due to
  broad parametric resonances. The bosons outnumbered the fermions
  during preheating as the Pauli exclusion principle prohibits
  explosive creation of fermions. As this process of creating
  particles via preheating is very rapid, the heavy bosons thus
  produced are initially far away from thermal equilibrium. But at a
  later stage when the particles initially thermalize during the
  generic reheating scenario then it may happen that few (or none) of
  the fermion species (and their partners) are initially thermalized
  and one obtains an over abundance of bosons over fermions. The other
  (or all) fermions and massless vector bosons (and possibly their
  partners) thermalize later when the Lee-Wick phase has evolved into
  the standard radiation dominated phase. In this case the energy
  density and the entropy density will predominantly have the heavy
  bosonic parts and their forms will be similar to the energy density
  and entropy density as given in Eq.~(\ref{energyd}) and
  Eq.~(\ref{entropyd}) where now $\tilde{g}_*$ and $\tilde{g}_{*s}$
  have to be replaced by their corresponding values which have
  predominantly heavy bosonic degrees of freedom. The thermodynamic
  evolution of such an universe matches with the one described in the
  beginning of the last section.

\item There is another possibility by which the constraint in
  Eq.~(\ref{ggrel}) can be satisfied and there can be high entropy
  generation at the time of reheating. If the Lee-Wick partners of the
  chiral fermions and the massless gauge bosons are much heavier and
  the lightest fermionic Lee-Wick partner's mass and the lightest
  Lee-wick massive vector boson (which is a partner of a standard
  gauge boson) mass are greater than the reheat temperature $T_{\rm
    Rh}$ then at reheating temperature the Lee-Wick partners of the
  fermions and gauge bosons must have decoupled from the plasma and
  consequently one must have only the heavy bosons with their Lee-Wick
  partners and standard fermions and gauge bosons contributing to the
  energy density of the universe. This scenario is favorable if one
  has a low reheat temperature $\sim 10^{6-5}\,{\rm GeV}$ or less. In
  this case the energy density and entropy density will have no
  contributions from the Lee-Wick partners of the, fermions and the
  massless gauge bosons, and the energy density and entropy density
  will be
\begin{eqnarray}
\rho=\frac{\tilde{M}^2}{24}\tilde{g}^{b}_{*N}T^2 + 
\frac{\pi^2}{30}\left( g_0^b + \frac{7}{8}\tilde{g}_F\right) T^4\,,
\label{energyd2n}
\end{eqnarray}
which is positive definite. The entropy will also turn out to be
positive definite. The factor $\tilde{g}^{b}_{*N}$ is exactly the same
as $\tilde{g}_{*N}$ except that it does not have the fermionic and
gauge boson part. The gauge bosons internal degrees of freedom are
encapsulated in the term $g_0^b$. The analysis of the thermodynamic
evolution of such an universe is qualitatively similar to the analysis
given in the previous section, specifically after Eq.~(\ref{mixede})
and Eq.~(\ref{mixeds}). As presently there is no theoretical model to
predict the range of masses of the Lee-Wick partner so the
assumption of having heavier fermionic and gauge boson partners
remains a possibility which can be tackled by future experiments.

Presently, the bounds on the electroweak Lee-Wick sector
\cite{Alvarez:2008ks, Underwood:2008cr, Alvarez:2008za} predict
Lee-Wick partners masses around $3\,{\rm TeV}$ or above. There is an
interesting observation regarding the production mechanism of Lee-Wick
fermion resonances \cite{Alvarez:2008za,Rizzo:2007ae}. The Lee-Wick
fermion resonances when produced will generally occur in pairs. Single
production of fermionic Lee-Wick resonances are suppressed by the
Lee-Wick fermions' mass. In a realistic situation this implies that
most of the heavy Lee-Wick fermion partners will not be in thermal
equilibrium if the reheat temperature is low. There will be more
single production of Lee-Wick bosonic resonances compared to fermionic
ones if the bosonic modes have masses comparable or lower than the
fermionic partner masses.
\end{enumerate}
 
The present model of Lee-Wick cosmology offers an early radiation
dominated universe whose energy density and entropy density can be
positive but they can be very small. If any of the above conditions
are fulfilled then one can circumvent this problem. The conditions
stated above are quite restrictive and consequently a Lee-Wick phase
in the early universe can definitely lead to a different kind of
cosmology as compared to the standard one.
\section{Discussion and conclusion}
\label{disc}
In standard radiation domination a usual particle can annihilate
itself by interacting with its antiparticle and producing photons in
thermal equilibrium. The entropy density remains constant and
consequently the temperature scales such that $Ta(t)$ remains constant
during this process of annihilation. In the Lee-Wick universe these
conventional scenarios do not hold any more. The simple reason why
they do not hold is related to the fact that the thermalized Lee-Wick
resonances have negative entropy density and consequently when these
thermal modes decouple they produce entropy. Entropy production
designates an out of equilibrium situation where the system is heated
up. As a result of this influx of energy the temperature of the system
also shoots up. The decoupling of any relativistic Lee-Wick thermal
resonance is an out of equilibrium process which will reheat the
universe. This reheating of the universe results in an increased
temperature of the universe.

From the nature of the Lee-Wick theories it was predicted that these
theories violate causality, as one has to apply future boundary
conditions to eradicate run away solutions \cite{Lee}. Lee and Wick
claimed and showed that violation of causality will not be observable
in low energy experiments. Application of the Lee-Wick paradigm in the
very early universe cosmology opens up a new question. It is a priory
not known how the accusal behavior of the Lee-Wick sector is going to
affect the causal history of the universe. The present authors do not
have an answer to this issue and it remains as a challenge to be
addressed in the future.

In the present article we initially present a toy model of Lee-Wick
thermodynamics where each standard boson or fermion has one Lee-Wick
partner respectively and more over the partners of the standard gauge
bosons have only two degrees of freedom. The qualitative features
coming out of this model predict the general nature of a Lee-Wick
resonance dominated early universe. In the realistic model one
considers two Lee-Wick fermion partners for each standard fermion
\cite{Wise, Espinosa} and also takes care of the longitudinal degree
of freedom of the massive Lee-Wick partners of the standard gauge
bosons. In the realistic Lee-Wick model it was shown that unless one
uses some specific assumptions about the bosonic and fermionic degrees
of freedom of the early universe it becomes difficult to produce
enough amount of energy density and entropy density during reheating.
One requires, in such a case, the bosonic degrees of freedom (arising
from the massive bosons) to outnumber the fermionic degrees of freedom
to achieve a workable model of the early universe. The general nature
of evolution in Lee-Wick cosmology remains the same as was presented
in the toy model under various limiting conditions. Each time a
Lee-Wick partner decouples from the cosmic plasma there will be a mini
reheating. The central idea of the present work is that if the
Lee-Wick partners are thermalized at reheating epoch after inflation
then there will be a series of explosive events when the temperature
and entropy of the universe will increase and consequently the system
will go out of thermal equilibrium. These events will happen at times
when Lee-Wick partners decouple from the cosmic plasma. The realistic
model of Lee-Wick theories which require two fermionic partners and
three degrees of freedom of the Lee-Wick partners of gauge bosons
demand a mass hierarchy of the partners where preferably the fermionic
partners are heavier or comparable in mass to their bosonic
counterparts if the reheat temperature is quite low ($T_{\rm
  Rh}\sim10^5$ GeV). In this case one has to assume that the mass of
the gauge boson partners to be quite high when compared to the
reheating temperature.  On the other hand, if the reheat temperature
is high ($T_{\rm Rh}\sim 10^{10}$ Gev), then a preheating scenario
\cite{Kofman}, which overproduces massive bosons than fermions and is followed
by reheating and thermalization of particles, is preferable to
describe an early universe dominated by Lee-Wick resonances.

\section*{Acknowledgements} 
We thank the anonymous referees for pointing out certain subtle issues
discussed in the article.


\end{document}